 \documentclass[journal,comsoc,twoside]{IEEEtran}

\usepackage{newtxtext,newtxmath}
\ifCLASSOPTIONcompsoc
  \usepackage[caption=false,font=footnotesize,labelfont=sf,textfont=sf]{subfig}
\else
  \usepackage[caption=false,font=footnotesize,labelfont=small]{subfig}
\fi
\ifCLASSOPTIONcaptionsoff
  \usepackage[nomarkers]{endfloat}
 \let\MYoriglatexcaption\caption
 \renewcommand{\caption}[2][\relax]{\MYoriglatexcaption[#2]{#2}}
\fi

\usepackage[utf8]{inputenc}
\usepackage{ragged2e}
\usepackage{units}
\usepackage{graphicx}
\usepackage{tabulary}
\usepackage{booktabs}
\usepackage{mathtools}  

\usepackage{amsmath} 
\usepackage{amsthm} 

\usepackage{amssymb}
\usepackage{amsfonts}
\usepackage[noadjust]{cite}
\usepackage{balance} 
\usepackage{tabularx}
\usepackage{comment}    
\usepackage{setspace}
\usepackage{xspace}
\usepackage{acronym}
\usepackage{multirow}
\usepackage{upgreek} 
\usepackage{nicefrac} 
\usepackage{ifthen}
\usepackage{csvsimple}
\usepackage{hyperref}
\usepackage{flexisym} 
\usepackage{fp} 
\usepackage{caption}
\usepackage{titlecaps}
\usepackage{optidef}
\usepackage[thinc]{esdiff} 
\theoremstyle{definition}

\usepackage{xargs} 
\usepackage{xcolor}
\usepackage{letltxmacro}

\LetLtxMacro{\oldsqrt}{\sqrt}
\renewcommand{\sqrt}[2][\mkern8mu]{\mkern-4mu\mathop{\oldsqrt[#1]{#2}}}

\usepackage{titlecaps}

\usepackage{graphicx}

\usepackage{acronym}
\acrodef{MFH}{Mobile Fronthauling}
\acrodef{PRA}{Prioritized Random Access}
\acrodef{ACK}{acknowledgement}
\acrodef{3GPP}{3rd Generation Partnership Project}
\acrodef{ACB}{Access Class Barring}
\acrodef{BSR}{Buffer Status Report}
\acrodef{BQA}{Bandwidth and QoS Aware}
\acrodef{CN}{Core Network}
\acrodef{CBR}{constant bit rate}
\acrodef{eNodeB}{evolved NodeB}
\acrodef{eNB}{evolved NodeB}
\acrodef{E-UTRAN}{Evolved Universal Terrestrial Radio Access Network}
\acrodef{FTP}{File Transfer Protocol}
\acrodef{GBR}{Guaranteed Bit Rate}
\acrodef{H2H}{human-to-human}
\acrodef{HARQ}{Hybrid Automatic Repeat Request}
\acrodef{HTC}{Human-type Communication}
\acrodef{IoT}{internet of things}
\acrodef{LTE}{Long Term Evolution}
\acrodef{LTE-A}{LTE-Advanced}
\acrodef{M2M}{Machine-to-Machine}
\acrodef{MTC}{Machine-type Communications}
\acrodef{PDCCH}{Packet Downlink Control Channel}
\acrodef{ePDCCH}{enhanced \acs{PDCCH}}
\acrodef{PDSCH}{Physical Downlink Shared Channel}
\acrodef{PUSCH}{Packet Uplink Shared Channel}
\acrodef{PLR}{packet loss ratio}
\acrodef{PRACH}{Physical Random Access Channel}
\acrodef{PDB}{Packet Delay Budget}
\acrodef{QCI}{QoS Class Identifier}
\acrodef{QoS}{quality of service}
\acrodef{RA}{Random-access}
\acrodef{RACH}{Random Access Channel}
\acrodef{RAN}{Radio Access Network}
\acrodef{RAR}{Random Access Response}
\acrodef{ROI}{RAN Overload Indicator}
\acrodef{RRC}{Radio Resource Control}
\acrodef{RRS}{RACH Resource Separation}
\acrodef{SR}{Scheduling Request}
\acrodef{UE}{User Equipment}
\acrodef{ML}{Machine Learning}
\acrodef{UL}{Uplink}
\acrodef{TA}{Time Alignment}
\acrodef{PUCCH}{Physical Uplink Control Channel}
\acrodef{PRB}{Physical Resource Block}
\acrodef{PCFICH}{Physical Control Format Indicator Channel}
\acrodef{PHICH}{Physical HARQ Indicator Channel}
\acrodef{DCI}{Downlink Control Information}
\acrodef{CCE}{Control Channel Element}
\acrodef{OFDM}{Orthogonal Frequency Division Multiplexing}
\acrodef{CFI}{Control Format Indicator}
\acrodef{OFDMA}{Orthogonal Frequency Division Multiple Access}
\acrodef{SC-FDMA}{Single-Carrier Frequency Division Multiple Access}
\acrodef{MME}{Mobility Management Entity}
\acrodef{S-GW}{Serving Gateway}
\acrodef{P-GW}{Packet Data Network Gateway}
\acrodef{TTI}{Transmission Time Interval}
\acrodef{nGBR}{non-GBR}
\acrodef{SDF}{Service Data Flow}
\acrodef{TFT}{Traffic Flow Template}
\acrodef{CP}{cyclic prefix}
\acrodef{FDD}{Frequency Division Duplexing}
\acrodef{TDD}{Time Division Duplexing}
\acrodef{EPS}{Evolved Packet System}
\acrodef{MCS}{Modulation and Coding Scheme}
\acrodef{DCCH}{Downlink Control Channels}
\acrodef{EPC}{Evolved Packet Core}
\acrodef{PER}{Packet Error Rate}
\acrodef{EAB}{Extended Access Barring}
\acrodef{TDPS}{Time-domain Packet Scheduling}
\acrodef{DL}{Downlink}
\acrodef{DSCH}{Downlink Shared Channel}
\acrodef{BLER}{Block Error Rate}
\acrodef{MNO}{mobile network operator}
\acrodef{PS}{Packet Scheduling}
\acrodef{TD}{Time-domain}
\acrodef{FD}{Frequency-domain}
\acrodef{FDPS}{Frequency-domain Packet Scheduling}
\acrodef{QPSK}{quadrature phase-shift keying}
\acrodef{RRM}{Radio Resource Management}
\acrodef{RAC}{Radio Admission Control}
\acrodef{RB}{Resource Block}
\acrodef{VoIP}{Voice-over-IP}
\acrodef{UCCH}{Uplink Control Channel}
\acrodef{HoL}{Head of the Line}
\acrodef{RLC}{Radio Link Control}
\acrodef{AL}{Aggregation Level}
\acrodef{SRS}{Sounding Reference Signal}
\acrodef{SPS}{Semi-Persistent Scheduling}
\acrodef{TPC}{Transmit Power Control}

\acrodef{3GPP}{Third Generation Partnership Project}
\acrodef{ACB}{Access Class Barring}
\acrodef{BSR}{Buffer Status Report}
\acrodef{CN}{Core Network}
\acrodef{eNodeB}{evolved NodeB}
\acrodef{GBR}{Guarantee Bit Rate}
\acrodef{H2H}{human-to-human}
\acrodef{HARQ}{Hybrid Automatic Repeat Request}
\acrodef{HTC}{Human Type Communication}
\acrodef{LTE}{Long Term Evolution}
\acrodef{LTE-A}{LTE-Advanced}
\acrodef{M2M}{Machine-to-Machine}
\acrodef{MAC}{Medium Access Control}
\acrodef{MTC}{Machine-Type Communications}
\acrodef{mMTC}{Massive Machine-Type Communications}
\acrodef{PRACH}{Physical Random Access Channel}
\acrodef{PDB}{Packet Delay Budget}
\acrodef{QCI}{QoS Class Identifier}
\acrodef{QoS}{quality of service}
\acrodef{RA}{Random Access}
\acrodef{RACH}{Random Access Channel}
\acrodef{RAN}{Radio Access Network}
\acrodef{RAR}{Random Access Response}
\acrodef{RAP}{Random Access Priorized}
\acrodef{ROI}{RAN Overload Indicator}
\acrodef{RAO}{Random Access Opportunity}
\acrodef{RRC}{Radio Resource Control}
\acrodef{RRS}{RACH Resource Separation}
\acrodef{SR}{Scheduling Request}
\acrodef{UE}{User Equipment}
\acrodef{UL}{Uplink}
\acrodef{TA}{Time Alignment}
\acrodef{PPA}{Preamble-Priority Aware}
\acrodef{PPA+}[ePPA]{Enhanced Preamble-Priority-Aware}
\acrodef{CAM}{Critical Alarm Messages}
\acrodef{RAP}{RA-Priorized}
\acrodef{LTA}{Lifetime-Aware}
\acrodef{GBR-LTA}{GBR-Priorized LTA}
\acrodef{CSS}{common search space}
\acrodef{DSS}{dedicated search space}
\acrodef{MPDCCH}{MTC \acs{PDCCH}}
\acrodef{UL-SCH}{Uplink Share Channel}
\acrodef{CR}{Contention Resolution}
\acrodef{LTE-Sim}{LTE Simulator}
\acrodef{FTTH}{Fiber to the Home}

\acrodef{ATM}{asynchronous transfer mode}
\acrodef{GEM}{GPON encapsulation method}
\acrodef{DBA}{dynamic bandwidth allocation}
\acrodef{EPON}{Ethernet PON}
\acrodef{ONU}{optical network unit}
\acrodef{ODN}{optical distribution network}
\acrodef{PON}{passive optical network}
\acrodef{SLA}{service level agreement}
\acrodef{VoIP}{voice over IP}
\acrodef{GPON}{Gigabit-capable \ac{PON}}
\acrodef{IPACT}{interleaved polling with adaptive cycle time}
\acrodef{DiffServ}{differentiated services}
\acrodef{OLT}{optical line terminator}
\acrodef{IEEE}{Institute of Electrical and Electronics Engineers}
\acrodef{TDM}{time division multiplexing}
\acrodef{TDMA}{time division multiple access}
\acrodef{MPCP}{multipoint control protocol}
\acrodef{WDM}{wavelength division multiplexing}
\acrodef{TWDM}{time and wavelength division multiplexing}
\acrodef{WDMA}{wavelength division multiple access}
\acrodef{TWDMA}{time and wavelength division multiple access}
\acrodef{TWMD}{time/wavelength division multiplexing}
\acrodef{FTTx}{fiber to the x}
\acrodef{PHY}{physical layer}
\acrodef{MAC}{media access control layer}
\acrodef{CoS}{class of services}
\acrodef{gGIANT}{group-GIANT}
\acrodef{XGPON}{10-Gigabit-capable PON}
\acrodef{GIANT}{Giga-PON access network}
\acrodef{T-CONT}{transmission container}
\acrodef{gT-CONT}{grouped T-CONT}
\acrodef{gAllocId}{group allocation ID}
\acrodef{XGEM}{GPON encapsulation mode}
\acrodef{EF}{expedited forwarding}
\acrodef{BE}{best effort}
\acrodef{AF}{assured forwarding}
\acrodef{BS}{Bandwidth Slicing}
\acrodef{mBS}{modified-BS}
\acrodef{IP}{interleaved polling}
\acrodef{DWBA}{dynamic wavelength and bandwidth allocation}
\acrodef{InP}{infrastructure provider}
\acrodef{TCO}{total cost of ownership}
\acrodef{CAPEX}{CAPital EXpenditure}
\acrodef{OPEX}{OPerating EXpenditure}
\acrodef{VNO}{virtual network operator}
\acrodef{MOS-IPACT}{IPACT with multi-ONU SLAs support}
\acrodef{SNS}{SubNetwork Slice}
\acrodef{MAC}{medium access control}
\acrodef{IP}{interleaved polling}
\acrodef{OLS}{onu load status}
\acrodef{RTT}{round to trip}
\acrodef{10G-EPON}{10 Gbit/s ethernet passive optical network}
\acrodef{X-GPON}{10 Gbit/s GPON}
\acrodef{ITU}{International Telecommunication Union}
\acrodef{DES}{Delayed Excess Scheduling}
\acrodef{GSF}{Grant scheduling frameworks}
\acrodef{GSP}{Grant sizing policy}
\acrodef{PEDB}{Policies for excess bandwidth distribution}
\acrodef{THR}{throughput}
\acrodef{FANS}{fixed access network sharing}
\acrodef{FE}{fair excess}
\acrodef{EC}{excess control}
\acrodef{HPS}{High Priority Subgroup First}
\acrodef{XGTC}{XG-PON transmission convergence}
\acrodef{NO}{network operator}

\acrodef{GSF}{grant scheduling framework}
\acrodef{GWP}{grant windows-size policy}
\acrodef{EDP}{excess distribution policy}
\acrodef{GSP}{grant scheduling policy}
\acrodef{TSF}{thread scheduling framework}
\acrodef{RTT}{round trip time}
\acrodef{InP}{Infrastructure service provider}
\acrodef{FL}{Federated Learning}
\acrodef{10G-EPON}{10 Gbit/s Ethernet Passive Optical Network}
\acrodef{3GPP}{3rd Generation Partnership Project}
\acrodef{3GPP}{Third Generation Partnership Project}
\acrodef{BAM}{Bandwidth-slicing with Adaptive-cycle and Multiple-wavelengths algorithm}
\acrodef{CNN}{Convolutional Neural Network}
\acrodef{ACB}{Access Class Barring}
\acrodef{ACK}{Acknowledgement}
\acrodef{AF}{Assured Forwarding}
\acrodef{Alloc-ID}{Allocation Identify}
\acrodef{APON}{ATM PON}
\acrodef{ASIC}{Application Specific Integrated Circuit}
\acrodef{ATM}{Asynchronous Transfer Mode}
\acrodef{AL}{Aggregation Level}

\acrodef{BE}{Best Effort}
\acrodef{BQA}{Bandwidth and QoS Aware}
\acrodef{BGP}{Bandwidth Guaranteed Polling}
\acrodef{BW map}{Bandwidth Mapping}
\acrodef{BLER}{Block Error Rate}
\acrodef{BSR}{Buffer Status Report}
\acrodef{BPON}{Broadband PON}

\acrodef{CAPEX}{Capital Expenditure}
\acrodef{CO}{Central Office}
\acrodef{CBR}{Constant Bit Rate}
\acrodef{CCE}{Control Channel Element}
\acrodef{CFI}{Control Format Indicator}
\acrodef{CN}{Core Network}
\acrodef{CR}{Contention Resolution}
\acrodef{CoS}{Class of Services}
\acrodef{CAM}{Critical Alarm Messages}
\acrodef{CP}{Cyclic Prefix}
\acrodef{CSS}{Common Search Space}

\acrodef{DBA}{Dynamic Bandwidth Allocation}
\acrodef{DBAM}{Dynamic Bandwidth Allocation with Multiple Services}
\acrodef{DSCH}{Downlink Shared Channel}
\acrodef{DCCH}{Downlink Control Channels}
\acrodef{DCI}{Downlink Control Information}
\acrodef{DES}{Delayed Excess Scheduling}
\acrodef{DSS}{Dedicated Search Space}
\acrodef{DiffServ}{Differentiated Services}
\acrodef{DL}{Downlink}
\acrodef{DPOA}{Dynamic Polling Order Arrangement}
\acrodef{DWBA}{Dynamic Wavelength and Bandwidth Allocation}

\acrodef{EC}{Excess Control}
\acrodef{EDP}{Excess Distribution Policy}
\acrodef{EF}{Expedited Forwarding}
\acrodef{EAB}{Extended Access Barring}
\acrodef{ePDCCH}{Enhanced \acs{PDCCH}}
\acrodef{eNodeB}{Evolved NodeB}
\acrodef{eNB}{Evolved NodeB}
\acrodef{E-UTRAN}{Evolved Universal Terrestrial Radio Access Network}
\acrodef{EPS}{Evolved Packet System}
\acrodef{EPC}{Evolved Packet Core}
\acrodef{EPDF}{Empirical Probability Density Function}
\acrodef{EPON}{Ethernet PON}

\acrodef{FE}{Fair Excess}
\acrodef{FPGA}{Field-programmable Gate Array}
\acrodef{FSD-SLA}{Fair Sharing with Dual SLAs}
\acrodef{FTP}{File Transfer Protocol}
\acrodef{FTTb}{Fiber To The Business}
\acrodef{FTTB}{Fiber To The Building}
\acrodef{FTTc}{Fiber To The cell}
\acrodef{FTTC}{Fiber To The Curb}
\acrodef{FTTO}{Fiber To The Office}
\acrodef{FTTH}{Fiber To The Home}
\acrodef{FTTdp}{Fiber To The Distribution Point}
\acrodef{FTTN}{Fiber To The Node}
\acrodef{FTTP}{Fiber To The Premises}
\acrodef{FTTx}{Fiber To The x}
\acrodef{FANS}{Fixed Access Network Sharing}
\acrodef{FD}{Frequency-domain}
\acrodef{FDPS}{Frequency-domain Packet Scheduling}
\acrodef{FDD}{Frequency Division Duplexing}

\acrodef{GEM}{GPON Encapsulation Method}
\acrodef{GIANT}{Giga-PON Access Network}
\acrodef{gAllocId}{Group Allocation ID}
\acrodef{gGIANT}{Group-GIANT}
\acrodef{GPON}{Gigabit Capable \ac{PON}}
\acrodef{GSF}{Grant Scheduling Framework}
\acrodef{GSP}{Grant Scheduling Policy}
\acrodef{gT-CONT}{Grouped T-CONT}
\acrodef{GTC}{GPON Transmission Convergence}
\acrodef{GWP}{Grant Windows-sizing Policy}

\acrodef{HGP}{Hybrid Granting Protocol}
\acrodef{HoL}{Head of the Line}
\acrodef{HPP}{High Priority Packets First}
\acrodef{HPS}{High Priority Subgroup First}
\acrodef{H2H}{Human-to-Human}
\acrodef{HTC}{Human Type Communication}
\acrodef{HARQ}{Hybrid Automatic Repeat Request}

\acrodef{IC}{Computing Institute}
\acrodef{IEEE}{Institute of Electrical and Electronics Engineers}
\acrodef{InP}{Infrastructure service Provider}
\acrodef{IoT}{Internet of Things}
\acrodef{IP}{Interleaved Polling}
\acrodef{IPACT}{Interleaved Polling with Adaptive Cycle Time}
\acrodef{ITU}{International Telecommunication Union}

\acrodef{LLID}{Logical Link ID}
\acrodef{LTA}{Lifetime-Aware}
\acrodef{LTE}{Long Term Evolution}
\acrodef{LTE-A}{LTE-Advanced}
\acrodef{LTE-Sim}{LTE Simulator}
\acrodef{LNF}{Largest Number of Frames First}
\acrodef{LRC}{Network Computer Laboratory}

\acrodef{mMTC}{Massive Machine-Type Communications}
\acrodef{M2M}{Machine-to-Machine}
\acrodef{MAC}{Medium Access Control}
\acrodef{MCS}{Modulation and Coding Scheme}
\acrodef{MNO}{Mobile Network Operator}
\acrodef{MME}{Mobility Management Entity}
\acrodef{MMF}{Multi-mode Optical Fiber}
\acrodef{MPCP}{Multipoint Control Protocol}
\acrodef{MTP}{Multi-thread Polling}
\acrodef{MTU}{Maximum Transmission Unit}
\acrodef{MOS-IPACT}{IPACT with Multi-ONU with SLAs Support}
\acrodef{MPDCCH}{MTC \acs{PDCCH}}
\acrodef{MTC}{Machine-Type Communications}

\acrodef{nGBR}{non-GBR}
\acrodef{NGA}{Next Generation Access Network}
\acrodef{NO}{Network Operator}

\acrodef{OAM}{Operation Administration and Maintenance}
\acrodef{ODN}{Optical Distribution Network}
\acrodef{OLS}{ONU Load Status}
\acrodef{OLT}{Optical Line Terminator}
\acrodef{OMCI}{ONU Management Control Interface}
\acrodef{OFDM}{Orthogonal Frequency Division Multiplexing}
\acrodef{OFDMA}{Orthogonal Frequency Division Multiple Access}
\acrodef{ONU}{Optical Network Unit}
\acrodef{OPEX}{Operating Expenditure}

\acrodef{P2MP}{Point-to-Multipoint}
\acrodef{P2P}{Point-to-Point}
\acrodef{PDCCH}{Packet Downlink Control Channel}
\acrodef{PDSCH}{Physical Downlink Shared Channel}
\acrodef{PEDB}{Policies for Excess Bandwidth Distribution}
\acrodef{PHY}{Physical Layer}
\acrodef{PUSCH}{Packet Uplink Shared Channel}
\acrodef{PLOAM}{Physical Layer OAM}
\acrodef{PLR}{Packet Loss Ratio}
\acrodef{PRA}{Prioritized Random Access}
\acrodef{PRACH}{Physical Random Access Channel}
\acrodef{PDB}{Packet Delay Budget}
\acrodef{PUCCH}{Physical Uplink Control Channel}
\acrodef{PRB}{Physical Resource Block}
\acrodef{PCFICH}{Physical Control Format Indicator Channel}
\acrodef{PHICH}{Physical HARQ Indicator Channel}
\acrodef{P-GW}{Packet Data Network Gateway}
\acrodef{PER}{Packet Error Rate}
\acrodef{PS}{Packet Scheduling}
\acrodef{PRACH}{Physical Random Access Channel}
\acrodef{PDB}{Packet Delay Budget}
\acrodef{PON}{Passive Optical Network}
\acrodef{PPA}{Preamble-Priority Aware}
\acrodef{PPA+}[ePPA]{Enhanced Preamble-Priority-Aware}

\acrodef{QCI}{QoS Class Identifier}
\acrodef{QoS}{Quality of Service}
\acrodef{QPSK}{Quadrature Phase-Shift Keying}

\acrodef{RA}{Random-Access}
\acrodef{RACH}{Random Access Channel}
\acrodef{RAN}{Radio Access Network}
\acrodef{RAR}{Random Access Response}
\acrodef{ROI}{RAN Overload Indicator}
\acrodef{RRC}{Radio Resource Control}
\acrodef{RRS}{RACH Resource Separation}
\acrodef{RRM}{Radio Resource Management}
\acrodef{RAC}{Radio Admission Control}
\acrodef{RB}{Resource Block}
\acrodef{RLC}{Radio Link Control}
\acrodef{RAP}{Random Access Priorized}
\acrodef{RAO}{Random Access Opportunity}
\acrodef{RTT}{Round-Trip Time}

\acrodef{SAR}{Smallest Available Report First}
\acrodef{SR}{Scheduling Request}
\acrodef{SC-FDMA}{Single-Carrier Frequency Division Multiple Access}
\acrodef{S-GW}{Serving Gateway}
\acrodef{SDF}{Service Data Flow}
\acrodef{SRS}{Sounding Reference Signal}
\acrodef{SPS}{Semi-Persistent Scheduling}
\acrodef{SR}{Scheduling Request}
\acrodef{SLA}{Service Level Agreement}
\acrodef{SNS}{SubNetwork Slice}
\acrodef{SMF}{Single-Mode Optical Fiber}
\acrodef{SNMP}{Simple Network Management Protocol}
\acrodef{SPD}{Shortest Propagation Delay First}
\acrodef{STP}{Single Thread Polling}

\acrodef{T-CONT}{Transmission Container}
\acrodef{TA}{Time Alignment}
\acrodef{TC}{Transmission Convergence}
\acrodef{TCO}{Total Cost of Ownership}
\acrodef{TD}{Time-Domain}
\acrodef{TDD}{Time Division Duplexing}
\acrodef{TDM}{Time Division Multiplexing}
\acrodef{TDMA}{Time Division Multiple Access}
\acrodef{TDPS}{Time-domain Packet Scheduling}
\acrodef{THR}{Throughput}
\acrodef{TFT}{Traffic Flow Template}
\acrodef{TLBA}{Two-Layer Bandwidth Allocation}
\acrodef{TTI}{Transmission Time Interval}
\acrodef{TPC}{Transmit Power Control}
\acrodef{TSF}{Thread Scheduling Framework}
\acrodef{TWDM}{Time and Wavelength Division Multiplexing}
\acrodef{TWMD}{Time-Wavelength Division Multiplexing}
\acrodef{TWDMA}{Time and Wavelength Division Multiple Access}

\acrodef{UCCH}{Uplink Control Channel}
\acrodef{UE}{User Equipment}
\acrodef{UL}{Uplink}
\acrodef{US}{Upstream}
\acrodef{DS}{Downstream}
\acrodef{US}{Upstream}
\acrodef{UL-SCH}{Uplink Share Channel}
\acrodef{mIPACT}{modified-IPACT}

\acrodef{MSD}{Multiple-Scheduling Domain}
\acrodef{SSD}{Single-Scheduling Domain}
\acrodef{WA}{Wavelength Agile}

\acrodef{WF}{Water-Fill}
\acrodef{FF}{First-Fit}
\acrodef{6G}{6th Generation mobile networks}
\acrodef{AI}{Artificial Intelligence}

\acrodef{VoIP}{Voice-over-IP}
\acrodef{VNO}{Virtual Network Operator}

\acrodef{WDM}{Wavelength Division Multiplexing}
\acrodef{WDMA}{Wavelength Division Multiple Access}

\acrodef{X-GPON}{10 Gbit/s GPON}
\acrodef{XGEM}{GPON Encapsulation Mode}
\acrodef{XGPON}{10 Gigabit Capable PON}
\acrodef{XGTC}{XG-PON Transmission Convergence}

\acrodef{IXR}{Immersive eXtended Reality}

\acrodef{CS}{Central Server}
\acrodef{FSL}{Federated Supervised Learning}
\acrodef{FRL}{Federated Reinforcement Learning}

\acrodef{FedAvg}{Federated Averaging}
\acrodef{KPI}{Key Performance Indicator}

\acrodef{CFS}{Centralized FL Server}
\acrodef{DFS}{Distributed FL Server}

\acrodef{LSTM}{Long Short-Term Memory Neural Network}

\newcommand{\etal}{\textit{et. al}\xspace}
\newcommand{\ie}{\textit{i.e., }\xspace}

\newcommand{\BS}{\textit{BS}\xspace}

\newcommand{\FLb}{\textit{FTS2}\xspace}

\ifCLASSOPTIONtwocolumn
  \newcommand{\soutc}[1]{
  }
  
\else
  \newcommand{\soutc}[1]{
  \textcolor{red}{{\sout{#1}}}
  }
  
\fi

\newcolumntype{L}[1]{>{\raggedright\let\newline\\\arraybackslash\hspace{0pt}}m{#1}}
\newcolumntype{C}[1]{>{\centering\let\newline\\\arraybackslash\hspace{0pt}}m{#1}}
\newcolumntype{R}[1]{>{\raggedleft\let\newline\\\arraybackslash\hspace{0pt}}m{#1}}

\newcommand{\issueone}{Delays of Federated Learning}
\newcommand{\issuetwo}{Large FL Packet Size}

\newcommand{\issuefour}{Bandwidth Slice for FL}

\newcommand{\policyone}{\textit{FL-first policy}\xspace}
\newcommand{\policytwo}{\textit{DC-first policy}\xspace}
\newcommand{\delaycritical}{delay-critical\xspace}
\newcommand{\computingtime}{computing time\xspace}
\newcommand{\BSx}{\textit{MW-BS}\xspace}
\newcommand{\FLx}{\textit{DWBA-FL}\xspace}

\title{Federated Learning over Next-Generation Ethernet Passive Optical Networks}

\author{
\IEEEauthorblockN{Oscar~J.~Ciceri, Carlos~A.~Astudillo, Zuqing Zhu, and Nelson~L.~S.~da~Fonseca}
\IEEEauthorblockA{Institute of Computing - University of Campinas \\
			    Campinas 13089-971, SP, Brazil \\
                Email: oscar@lrc.ic.unicamp.br, castudillo@lrc.ic.unicamp.br,
                zqzhu@ieee.org,
                nfonseca@ic.unicamp.br}
}

\begin{document}

\author{
\IEEEauthorblockN{Oscar J. Ciceri, Carlos~A.~Astudillo, Zuqing Zhu, and Nelson~L.~S.~da~Fonseca%
\thanks{
}%
\thanks{O.~J.~Ciceri, C.~A.~Astudillo and N.~L.~S~da~Fonseca are with the Institute of Computing, University of Campinas 13083-852, Brazil (emails: oscar@lrc.ic.unicamp.br,castudillo@lrc.ic.unicamp.br, nfonseca@ic.unicamp.br).}%
\thanks{Z.~Zhu is with the School of Information Science and
Technology, University of Science and Technology of China, Hefei, Anhui 230027, P. R. China (email: zqzhu@ieee.org).}%
} }

\maketitle
\begin{abstract}
Federated Learning (FL) is a 
distributed machine learning (ML) type of processing that preserves the privacy of user data, sharing only the parameters of ML models with a common server.  
The processing of FL requires specific  latency and bandwidth demands  that need to be fulfilled by the operation of the communication network.   
%
This paper introduces a Dynamic Wavelength and Bandwidth Allocation algorithm for Quality of Service (QoS) provisioning for FL traffic over 50 Gb/s Ethernet Passive Optical Networks. 
The proposed algorithm prioritizes FL traffic and reduces the delay of FL and delay-critical applications supported on the same infrastructure.
%
%

\end{abstract}


	\section{Introduction}
\label{sec:introduction}
\acresetall

\ac{AI} applications are widely employed in various different sectors of our daily life, such 
as in smart cities, smart homes, smart transportation, smart health, and finance. 
Moreover, the upcoming \ac{6G} will support an even wider  range of applications based on \ac{AI}, such as holographic communications, super-smart homes, and cooperative autonomous robots. 
%
The \ac{6G} system will change the network vision from  "connected things"  to "connected intelligence" \cite{8808168}, in which advanced and ubiquitous \ac{AI} will empower numerous   applications. 
%
%

Traditional use of \ac{ML} models relies on batch processing in a central server and the employment of datasets containing user  data. With the worldwide adoption of data protection and privacy legislation, the creation of datasets  and  applications based on ML has been considerably limited. One way of coping with such restrictions is the adoption of \acf{FL}, which is a distributed way of processing machine learning algorithms that does not disclose  private data. In FL, clients train a local ML model using a private dataset, and the parameters of these local models are then sent to a central server. The server produces a global model on the basis of the numerous parameter values received and distributes this global model to the clients for further training. This round of processing is  repeated until the global model produces results with an acceptable level of accuracy.  In this way, user privacy is preserved. The most common approaches for the consolidation of the parameters sent by the clients  to produce the global model (FedAvg) rely on the assumption that clients are  synchronized  and that local datasets are independent and identically distributed \cite{mcmahan2017communication}. Such  FL can be used for the training of large ML models involving thousands or even millions of parameters. 

The processing of \acl{FL} models has brought challenges to  communication networks.
 \ac{FL} clients may produce highly bursty traffic when uploading their  model parameters  to the server.
For example,    clients training a   \acp{CNN} model with  few convolution layers and thousands of parameters may need to send   hundreds of  megabytes. When millions of parameters are involved, the amount of bytes sent can be of the order of  gigabytes    \cite{Kaiming_CVPR2016-7780459}.
%
Moreover, \ac{FL} may impose stringent communication delays  for the uploading  of client parameters to enhance   fast convergence to the global model, especially when the federation involves numerous clients. 
To cope with diverse communication delays, the server may either wait the arrival of the local  parameters from all clients,  increasing convergence time, or exclude the late arriving data from the  parameter consolidation step, which reduces the  accuracy of the model \cite{li2020bandwidth}.
%
In addition, \ac{FL} may also require a very large number of training rounds to produce accurate global models \cite{caldas2018leaf}. These challenges calls for efficient  resource allocation mechanisms to meet the FL requirements.

\acf{PON} is a cost-efficient access network technology for delivering broadband services \cite{flavio}.
%
%
Operators have already deployed 10 Gb/s \ac{TDM} PONs during the past two decades. In recent years, the ITU and IEEE standardization groups have proposed  next-generation \acp{PON} based on \ac{TWDM} to increase the network capacity for supporting  demanding applications and services. TWDM allows allocation in various wavelength channels of 25 Gb/s (50G-EPON) and 10 Gb/s  (40G-XPON) \cite{wey2020outlook}.
%
%

\acp{InP} owning a \ac{PON} can increase network utilization, as well as their  profits, by offering a diverse variety of services  to a variety of different customers, such as residential users, enterprises, and mobile network operators (MNO). The capacity to support the Quality of Service (QoS) requirements of emerging 5G/6G applications, such as 
%
distributed \acl{ML}, Tactile Internet, or \ac{MFH} calls for efficient use of network resources.
%
%
Nonetheless, the aforementioned challenges of \ac{FL} applications make QoS provisioning challenging.

A few approaches have been proposed to deal with FL processing 
over PONs  (\cite{li2020bandwidth,li2020scalable}). 
One of these is the  \ac{BS} approach  for TDM-PONs, that reserves  portions of the PON capacity for \ac{FL} clients  \cite{li2020bandwidth}. The bandwidth for each of the slices granted per cycle is orders of magnitude less than that required for transmitting a model update, 
which  implies that several scheduling cycles will be required to fully upload the   parameters of the client models.  
An architecture for scalable federated learning involving two-step of aggregation over \acsp{PON} was introduced in \cite{li2020scalable}. 
In this proposal, the parameters of local models are first aggregated at clients connected to \ac{ONU}  and then aggregated on a server connected to the \ac{OLT}.
As a consequence, the amount of  upstream traffic remains relatively constant regardless of the number of clients in the federation.
However, the high volume of traffic generated by the FL clients can create network bottlenecks, which impacts on    the time to upload of the local parameters.

%

%
This paper discusses the main issues for supporting FL applications over PONs and introduces a \ac{DWBA} algorithm for 50G-EPONs that dynamically prioritizes  \ac{FL} traffic while maintaining the traditional guaranteed bandwidth scheme for  \ac{PON} customers.
Two prioritization policies have been proposed to reduce the delay of \ac{FL} traffic and \delaycritical applications. 
In the first, the intra-ONU scheduler strictly prioritizes the \ac{FL} traffic over that from  other  types of applications (\policyone). 
In the second, the \delaycritical traffic is prioritized over  the FL traffic (\policytwo).
The \ac{BS} algorithm introduced by Li \etal (2020) \cite{li2020bandwidth} was adapted to employ multiple wavelengths, as well as   an adaptive polling cycle as required in 50G-EPON networks with dynamic resource allocation for comparison purpose.

Results show that the \policytwo increases the \ac{FL} model accuracy and reduces the delay of federated learning  and \delaycritical applications when compared to the \ac{BS} approach and the \policyone.

	\section{Resource Allocation in Passive Optical Networks}
\label{sec:PON_standards}

\ac{PON} is a  network access technology that offers larger  capacity, greater cost-efficiency, and more energy savings than do other network access technologies.
There are two main \ac{PON} standards: \ac{EPON} and \ac{GPON}, with \ac{EPON} being  less expensive.
 \ac{GPON}  transmission system ofemploys  synchronous frames  issued at every \unit[125]{$\mu$s}, while those of \ac{EPON} use  Ethernet frames  asynchronously for transmissions based on granted cycles of variable duration. 
While traditional \ac{PON} standards allow bit rates of \unitfrac[1]{Gb}{s} and \unitfrac[10]{Gb}{s},
the next-generation \acs{PON} standards allow those of  \unitfrac[40]{Gb}{s} to \unitfrac[100]{Gb}{s}.


The 50 Gb/s optical access network standardized in \acs{IEEE} 50G-EPON  802.3ca-2020 \cite{9135000} is a promising technology for adoption  by  \ac{InP} to support emerging services with strigent latency and bandwidth requirements.
%
%
%
%
%
This 50G-EPON technology employs the \ac{TWDMA} technique for controlling  uplink transmissions between the \acp{ONU} and the \ac{OLT}.
There are three main \acs{TWDM}-\acs{PON}-based access architectures for the connectivity between the \ac{OLT} and \acp{ONU}: \ac{MSD}, \ac{SSD}, and \ac{WA}. 
In the first,  ONUs transmit on a single wavelength at a time. In the second,  ONUs can transmit simultaneously on all wavelengths, and in  the third, more than one wavelength can be granted to a single \ac{ONU}. 

In this technology, the signaling protocol \ac{MPCP}  is employed for resource allocation. 
This protocol uses  Report and Gate  messages for this  allocation. 
Report messages are sent on the upstream to the \ac{OLT} by the \acp{ONU}  to request bandwidth, while  
 Gate messages  are sent on the downstream by the \ac{OLT} to the \acp{ONU} to inform the grated  wavelength(s) and transmission windows, as well as  the starting time of the next transmission window. 
Resource allocation is  carried out in two steps, one for wavelength allocation and the other for bandwidth  allocation. The use of different schemes for transmission on multiple wavelengths can be defined  on the basis of 
conventional \ac{DBA}  algorithms for TDM-PONs.


For dynamic bandwidth allocation over EPONs, the \ac{IPACT} algorithm has been adopted to complement the \ac{MPCP} protocol.
The \ac{IPACT} algorithm  employs an interleaved polling and statistical multiplexing technique that leads to efficient upstream channel usage.
The limited policy has been used to assure bandwidth to \acp{ONU} according to pre-defined Service Level Agreements.
Moreover, the original \ac{IPACT} algorithm employs a single wavelength channel for scheduling. It has been modified to operate with multiple wavelengths in \cite{mcgarry2006wdm}, \cite{wang2017dynamic} and \cite{hussain2017low}.  
The modified \ac{IPACT} algorithm was  proposed for the \ac{SSD} and \ac{MSD} architectures \cite{mcgarry2006wdm}.
Additional algorithms have been proposed: the \ac{WF} \cite{wang2017dynamic}   to promote fairness in the wavelength utilization and \ac{FF} \cite{hussain2017low} to provide less delay. 
Moreover, when there is no scheduler for  \ac{QoS} provisioning in the \ac{PON}, the First-Come-First-Served (FCFS) queuing policy is employed. However, this strategy does not consider the priority or required bandwidth/delay of the applications.

\begin{figure*}[!t]
    \centering
    \includegraphics[width=0.7\linewidth]{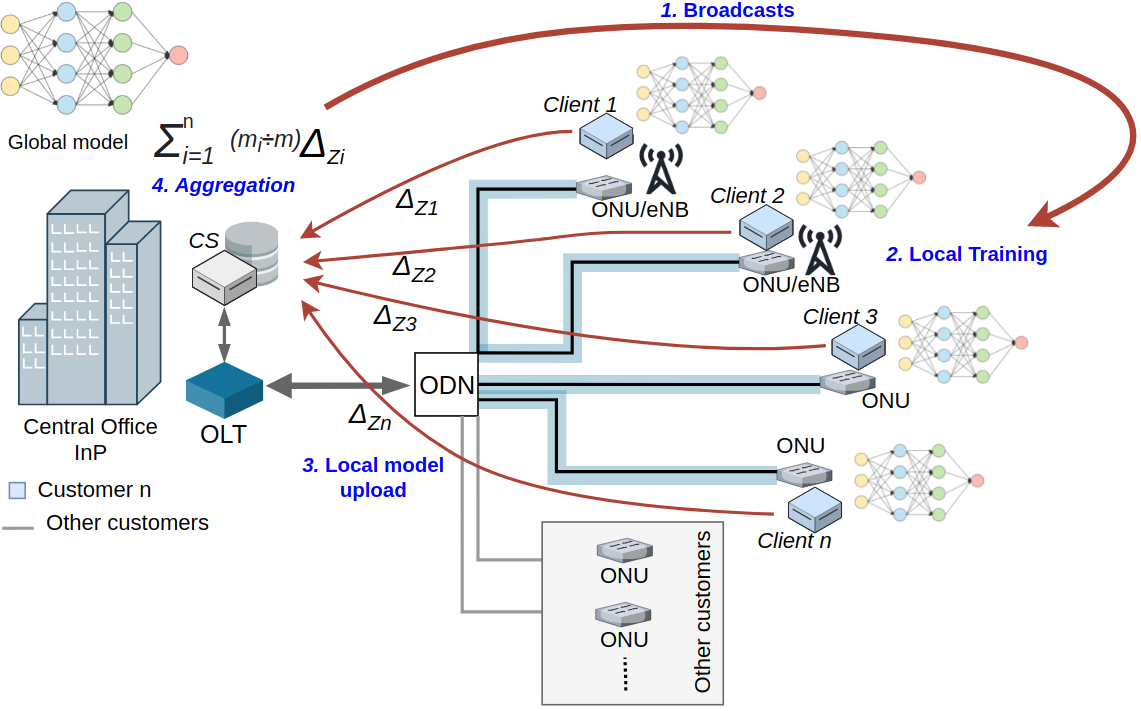}
     \caption{An overview of federated learning over passive optical networks.}
    \label{fig:arq}
\end{figure*}

The performance of  diverse applications in a \ac{PON} is ensured by  the adopted \ac{QoS} mechanism, it controls the way  frames are queued, prioritized, and scheduled.
Such assurance of \ac{QoS} can be provided by either the \ac{ONU} or \ac{OLT}.
In the single-level architecture, the \acp{ONU} reports  individual queue sizes, while the \ac{OLT} distributes the bandwidth for each type of traffic.
In the hierarchical architecture, the \ac{OLT} allocates bandwidth for each \ac{ONU}, and the \acp{ONU} manage the amount of  bandwidth to be allocated for each traffic.

The most straightforward method of facilitating \ac{QoS} provisioning is the Differentiated Service approach. It  classifies network traffic and delivers different services to different applications.
The simplest way to implement  Differentiated Services is to employ  strict priority scheduling. The ONUs categorize the incoming traffic and put it in the buffer, imposing the   prioritization of different  traffic classes.
%
%
With the employment of Differentiated Services, however, the \ac{PON} can support packetized voice and video with strict bandwidth and  latency constraints,  as well as best-effort  traffic \cite{mcgarry2008ethernet}.
However, none of the exiting mechanisms have been specially designed to support the \ac{QoS} requirements of  \ac{FL} applications.


\section{Resource Allocation for  Federated Learning }
\label{sec:FL_Issues}

Fig. \ref{fig:arq} illustrates the scenario of FL processing over a PON; where clients are connected to the ONUs and the server is attached to the OLT.

The FL training process can be either asynchronous or synchronous.
In the former, the global model parameters are computed as soon as the server receives updates of the parameters of the local models from a certain number of clients. 
%
%
In the latter, the server aggregates the local parameters  that arrive in a period of constant duration
excluding the parameters from the late arriving stragglers. This exclusion, however, reduces the  accuracy of the model as well as increases the time required for  convergence to a final global model.

\begin{figure}
\centering
\subfloat[\titlecap{\issueone}]{\label{fig:allocation_problem}%
\includegraphics[height=2.3in]{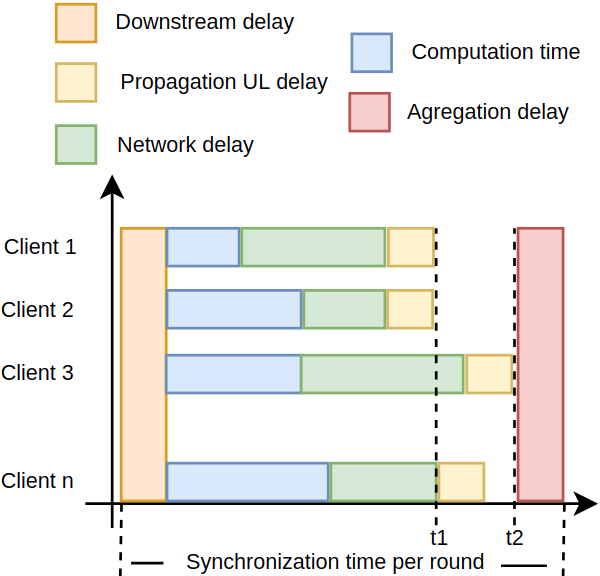}}%
\qquad
\subfloat[\titlecap{\issuetwo}]{\label{fig:size_problem}
\includegraphics[height=2.3in]{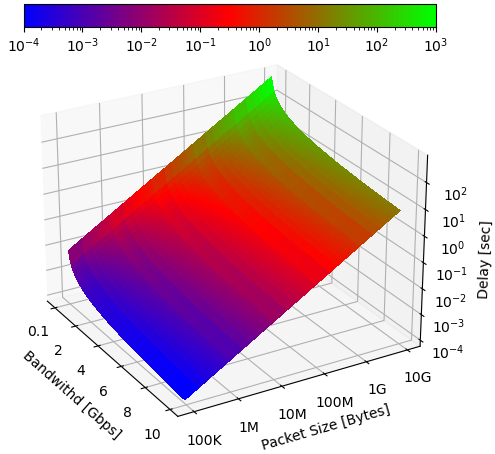}}%
\qquad
\subfloat[\titlecap{\issuefour}]{\label{fig:BS_FL}%
\includegraphics[height=2.3in]{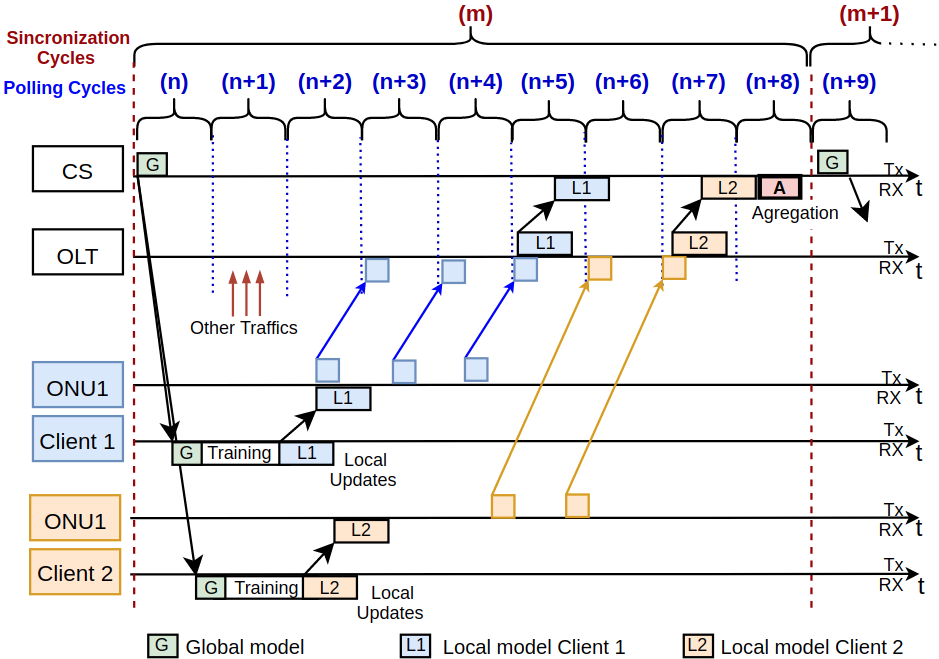}}%
\caption{Resource allocation issues for \acl{FL} over PONs}
\label{fig:issues}
\end{figure}

The synchronization time per round includes the downstream, computing, network, propagation, and aggregation delays, as shown in Fig. \ref{fig:allocation_problem}. 
The downstream delay includes the propagation and transmission delays of the parameters of the global model from the server to the clients. 
The \computingtime is the time taken to train  the local model at the client in each round. 
The network delay is the  time spent in communication the local model parameters from the clients to the server.
The propagation delay is the time taken by the bits transmitted  to travel from the client to the server on the network medium.
The aggregation delay is the processing time of the aggregation algorithm. 
%


The network delay depends  on  the network load and the resource allocation mechanism employed for the \ac{PON} for allocating bandwidth and wavelength(s) to the ONUs. The \computingtime depends on the  capacity  of the client and the  size of the training dataset, while the propagation delay depends on  the distances between the server and the clients.
Moreover,  long network delays may increase the number of straggler clients,  decreasing the model accuracy and increasing the convergence time.
The time taken to transmit the local model parameters to the server depends on the bandwidth allocated to the \ac{FL} traffic. 
In general, \ac{PON} customers receive a portion of the total available bandwidth in the \ac{PON} due to the shared nature of the upstream channel.
%
Residential and business customers usually have  guaranteed bandwidth from tens to hundreds of Mbps, and other PON customers can require on demand up to tens of Gbps.
%
However, the large size of the local model parameters, which can be in the order of gigabytes, may demand several seconds to be fully transmitted, even with guaranteed bandwidths on the order of Gbps, as shown in Figure \ref{fig:size_problem}.

The unique characteristics of  \ac{FL} applications introduce challenges for the management of the available bandwidth in scenarios with limited bandwidth and diversity of customers, services and applications.
%
%
The recently proposed  \ac{BS} algorithm  to the support of \ac{QoS} for \ac{FL} applications assures a slice of bandwidth for the \ac{FL} traffic. This is then allocated according to the ascending order of  downstream client delay and \computingtime. 
However, the approach is not efficient for \ac{PON} scenarios when loads are high and traffic is bursty.
The large bandwidth slice  required to serve  \ac{FL} traffic properly
 reduces the available bandwidth for other \ac{PON} customers. It also reduces the statistical multiplexing gain due to the static allocation of the network resources.
On the other hand, if the slice is much smaller than the total available bandwidth, the granted bandwidth is likely to be insufficient to serve timely the FL processing due to the large size of the  \ac{FL} packets. 
As a consequence, clients send a small portion of the \ac{FL} frames per cycle, requiring numerous scheduling cycles to be fully served.
If two or more clients have \ac{FL} frames in the queue at a given time, then only one client can use the slice, while the others will have to wait for that slice to become available, as shown in Figure~\ref{fig:BS_FL}.

Furthermore, the \ac{BS} approach is not compatible with  traditional \ac{PON} business models, in which customers rent  portions of the \ac{PON} capacity from the \ac{InP} to support their applications and services.   
Moreover, it is unclear who pays for a shared bandwidth slice. 

In summary, even though the \ac{BS} approach reduces the latency for \ac{FL} applications in relation to the traditional First Come First Served approach, the static bandwidth allocation and the incompatibility of the approach with traditional business models may  lead to issue of \ac{QoS} support and deployment.

%



%

%

%

	\section{\ac{DWBA} Schemes for Supporting \ac{FL} traffic over 50G-\ac{EPON} networks}
\label{sec:proposal}

This section introduces the proposed scheme for providing  \ac{QoS} for \ac{FL} applications 
over 50G-EPON networks. 

\subsection{Adaptation of the Bandwidth Slicing Approach to \mbox{TWDM-EPON}}

The \ac{BS} algorithm \cite{li2020bandwidth} serves FL traffic by dynamically allocating bandwidth on a single wavelength for  FL clients. It calculates the number of cycles an FL client requires to be completely served based on the required bandwidth and the fixed polling cycle length of \unit[125]{$\mu s$} employed in the GPON technology. 
However, next-generation PONs employ multiple wavelengths and the cycle duration is  unknown a priori when an adaptive polling cycle mechanism is employed, such as in  the EPON technology.
%
Therefore, the proposed \ac{BS} algorithm cannot be directly employed in TWDM-EPON networks.

We proposed an adaptation of the \ac{BS} approach for TWDM-EPONs called multi-wavelength BS algorithm (\BSx). It deals with  multiple wavelengths and employs an adaptive polling cycle for dynamic resource allocation. 
A portion of the PON capacity (bandwidth slice) is still reserved for the \ac{FL} traffic in each scheduling cycle, but
%
instead of using the polling cycle information to grant the transmission windows for FL traffic and then share the remaining slice capacity with other traffic types, \BSx  reserves the total  slice  to the FL traffic as long as a bandwidth request from any FL client exists. The use of a dynamic polling cycle  reduces the FL traffic delays and avoids the need for information about the duration of the unknown  upcoming cycles.
%
%
Three variations of the MW-BS algorithm were proposed for different 
\ac{TWDM} wavelength allocation policies,
namely \BSx-\textit{SSD}, 
\BSx-\textit{MSD}, and
\BSx-\textit{FF}.

\begin{figure*}[!t]
    \centering
    \includegraphics[width=0.7\linewidth]{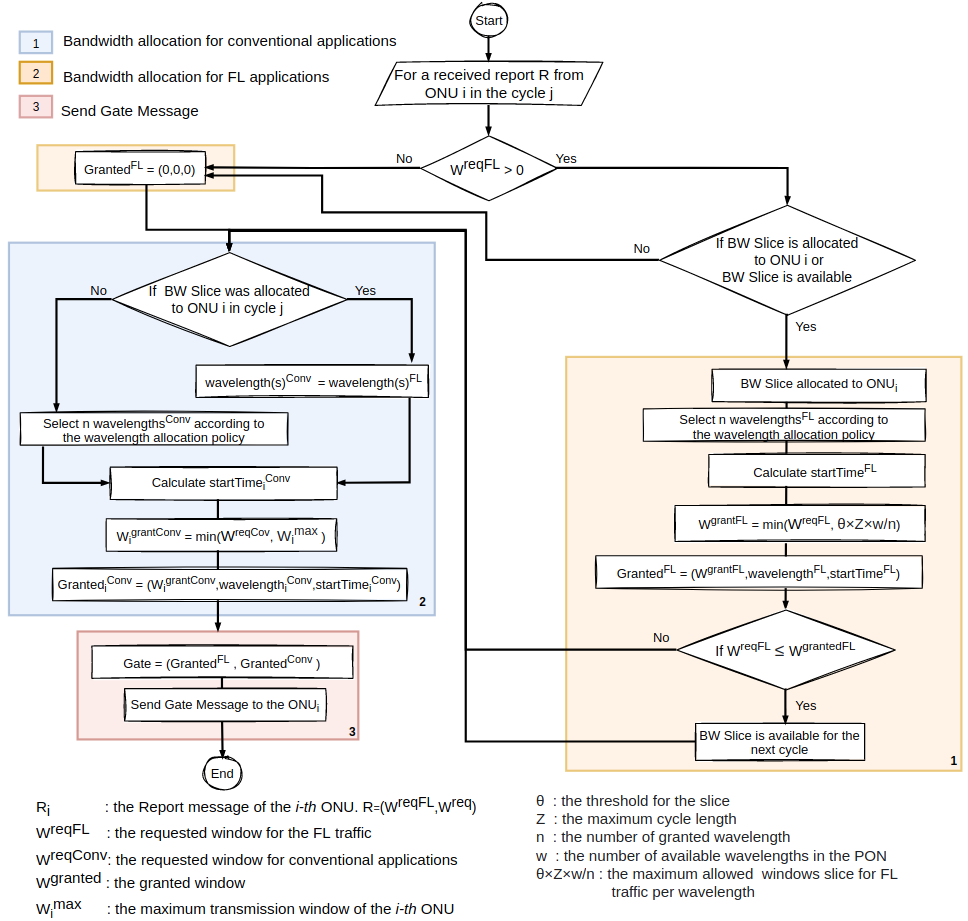}
     \caption{Flow chart of the bandwidth slicing algorithm adapted for TWDM-EPONs.}
    \label{fig:alg}
\end{figure*}

The flow diagram in Figure \ref{fig:alg} summarizes the proposed DWBA scheme residing on the OLT.
The \acp{ONU} sends the Report message requesting bandwidth for  \acl{FL} as well as  conventional applications.
When a Report message arrives from a \ac{ONU}
containing a bandwidth request for FL traffic, the OLT first grants the bandwidth from the reserved slice, if  available (Block 1); 
otherwise, the \ac{OLT} allocates bandwidth for the conventional traffic (Block 2).

For allocation of  bandwidth for the slice,
The \ac{OLT} also reserves  bandwidth  for the \ac{ONU} for  upcoming cycles. 
It selects the wavelength(s) as a function of the
%
\ac{TWDM} wavelength allocation policy,
and calculates the next starting time for the \ac{FL} transmission. 
For the SSD policy, the \ac{OLT} grants all wavelengths. 
For the MSD policy, the \ac{OLT} grants a predetermined-fixed wavelength. 
For the FF policy, the \ac{OLT} grants the first available  wavelength, 
and then calculates the granted transmission window for each allocated wavelength, depending on the number of wavelengths and the portion of the \ac{PON} capacity designed for \ac{FL} use.
If the granted window is equal to the requested windows, the \ac{FL} traffic will be fully served, and the \ac{OLT} will make the bandwidth slice available for the next cycle.

The \ac{OLT} also calculates the granted bandwidth for the convectional applications. 
If  the \ac{OLT} has previously allocated the  bandwidth for the slice, it selects these  wavelength(s)  for the \ac{FL} traffic.
Otherwise, the \ac{OLT} selects the wavelength(s) for the FL traffic and calculates the next starting time for that \ac{FL} transmission as a function of  the \ac{TWDM} wavelength allocation policies involved. The transmission window for the conventional applications is calculated  according to the limited policy.
Finally, the \ac{OLT} issues and send a Gate message with the granted bandwidths for both \ac{FL} and conventional  applications.

\subsection{DWBA for Federated Learning}
To address the issues of resource allocation  raised here, we introduce a DWBA algorithm that supports \ac{QoS} provisioning for \acl{FL} applications while meeting the requirements of \delaycritical applications in TWDM-\ac{EPON} networks. The algorithm is called DWBA for Federated Learning (DWBA-FL).

The idea behind our proposal is to allow  \ac{PON} customers to employ their guaranteed bandwidth for the   scheduling of the \ac{FL} application,  without jeopardizing the QoS provisioning of other \delaycritical applications. To achieve this, the proposed mechanism adopts the widely-used Differentiated Service approach to tackle the \ac{QoS} provisioning problem of \acl{FL} applications over Ethernet \ac{PON}. The prioritization of FL traffic is used to comply with the traditional business model, as well as to improve statistical multiplexing gain.
%
%

The proposal employs an intra-\ac{ONU} scheduler with a strict priority queuing policy to serve the ONU queues.  
The \acp{ONU} arbitrates the transmission demands of the different applications.
Upon the arrival of  a Report message, the \ac{OLT} calculates the transmission window according to the conventional limited policy and selects the  wavelength(s)  on the basis of the 
%
\ac{TWDM} wavelength allocation policies.
The \ac{OLT} then  sends a Gate message containing the resource allocation decision.  
Upon the receipt of that Gate message, the intra-ONU scheduler distributes the received bandwidth among the  queues in the ONU. 
In our model,  traffic is classified as \acl{FL}, \delaycritical, delay-sensitive, or best-effort.
The \acp{ONU} mantains four different queues for buffering frames for these types of traffic.

We propose two prioritization policies.
The \policyone defines the \ac{FL} traffic as that of the $highest$ priority and the delay-critical, delay-sensitive, and best-effort  being  of  $high$, $medium$ and $low$ priority, respectively. 
On the one hand, this strict prioritization of \ac{FL} frames can reduce synchronization time for  \ac{FL} processing. 
It can also  increase the delay of  \delaycritical application because the \ac{FL} traffic requires a large bandwidth per cycle.
To help alleviate this problem, we propose the \policytwo, in which delay-critical traffic also has the highest priority and \acl{FL} traffic the $high$ priority.
Moreover, the proposal was defined for all  \ac{TWDM} architectures.

	\section{Performance Evaluation}
\label{sec:performance-evaluation}

The performance of the proposed \ac{DWBA} scheme was evaluated  using an \ac{EPON} simulator (EPON-Sim), previously validated in \cite{ciceri2021PON_5G_MFH}. 
This extension was extended to support the three architectures, \ac{SSD}, \ac{MSD} and \ac{WA},  proposed for 50G-EPON networks.
Moreover,  our proposal and the  \ac{BS} approach were introduced in the simulator.
%
%
 
%


\subsection{Simulation Model and Setup}
\label{sec:simulation-model}

The simulation scenarios include a  50G-\ac{EPON} network with $1$ \ac{OLT} serving 32 \acp{ONU} on an optical distribution network with a tree topology. 
Two wavelength channels of 25 Gbit/s were employed for upstream transmission, giving a total capacity of 50 Gbit/s.
The total available bandwidth in the \ac{PON} was equally distributed among the \acp{ONU}, 
so that each \ac{ONU} has the same guaranteed bandwidth $b_i$, while the aggregated offered load per \ac{ONU} $l_i$ varied from $0.6\cdot b_{i}$ to $1.0\cdot b_{i}$ (for the sake of clearness and brevity, herein after, $b_{i}$ is omitted from the offered load values of ONU $i$).

The aggregated load included the traffic of the four different types of application: \acl{FL}, \delaycritical, delay-sensitive, and \acl{BE}.
The benchmarking framework for learning in federated settings LEAF \cite{caldas2018leaf}  was used to generate the \ac{FL} traffic. 
The FEMNIST dataset and  \ac{CNN} with two 5×5 convolution layers were used for model training, while the FedAvg algorithm was employed to aggregate the local parameters in the server.
%
%
Other configurations for the learning process, such as learning rate and batch size, followed the settings defined in  \cite{li2020scalablex}. 
 \ac{FL} clients generated  \unit[26.4]{MBytes} of data in each round of training. 
Moreover, the \acp{ONU} put the local parameters into frames according to the Ethernet protocol, which has a Maximum Transmission Unit of \unit[1500]{bytes} and a header field for signaling (preamble)  of \unit[20]{bytes}.
%
%

The \delaycritical applications were modeled employing a \ac{CBR} flow. It was coded with a fixed-size packet of \unit[70]{bytes} and an inter-arrival time of \unit[12.5]{$\mu$s}, which produces an offered load of \unit[44.8]{Mbps}.
The rest of the offered load $l_i$ was evenly distributed between delay-sensitive and \acl{BE} traffic. 
The traffic streams were generated employing Pareto ON-OFF sources. 
The ON period time and packet-burst size followed a Pareto and Bounded  Pareto distributions, respectively. The aggregated traffic at the ONU had a  Hurst parameter of 0.8.
Moreover, the packet lengths are uniformly distributed between $64$ and $1518$ bytes.

A threshold value of $\theta=0.015$  was employed in the \BSx algorithm, as in \cite{li2020bandwidth}.
This algorithm reduces the bandwidth  for each \ac{ONU} since it reserves bandwidth  for the slice.
Moreover, we employ the same aggregated offered load in the simulated algorithms to make  a fair comparison.
The duration of the guard period was set to \unit[0.624]{$\mu s$} with a  maximum cycle length of \unit[1]{ms}.
Each simulation scenario lasted \unit[100]{s} and was replicated $10$ times. 

%


\begin{figure*}
\centering
\subfloat[Delay-critical application]{\label{fig:ef}
\includegraphics[height=2.8in]{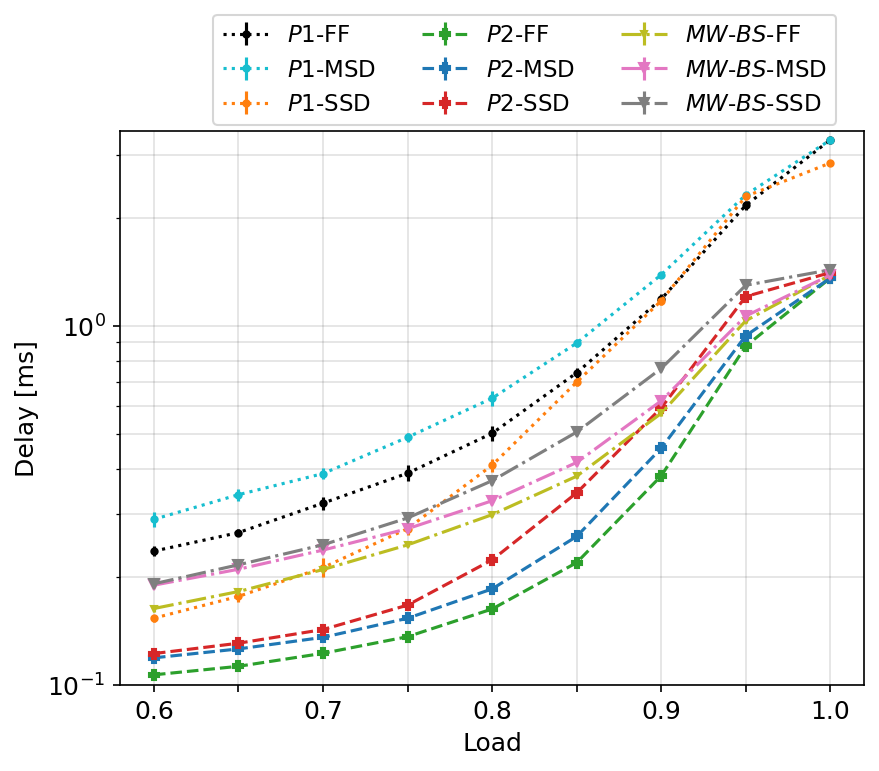}}%
\qquad
\subfloat[\acl{FL} application]{\label{fig:fl}%
\includegraphics[height=2.8in]{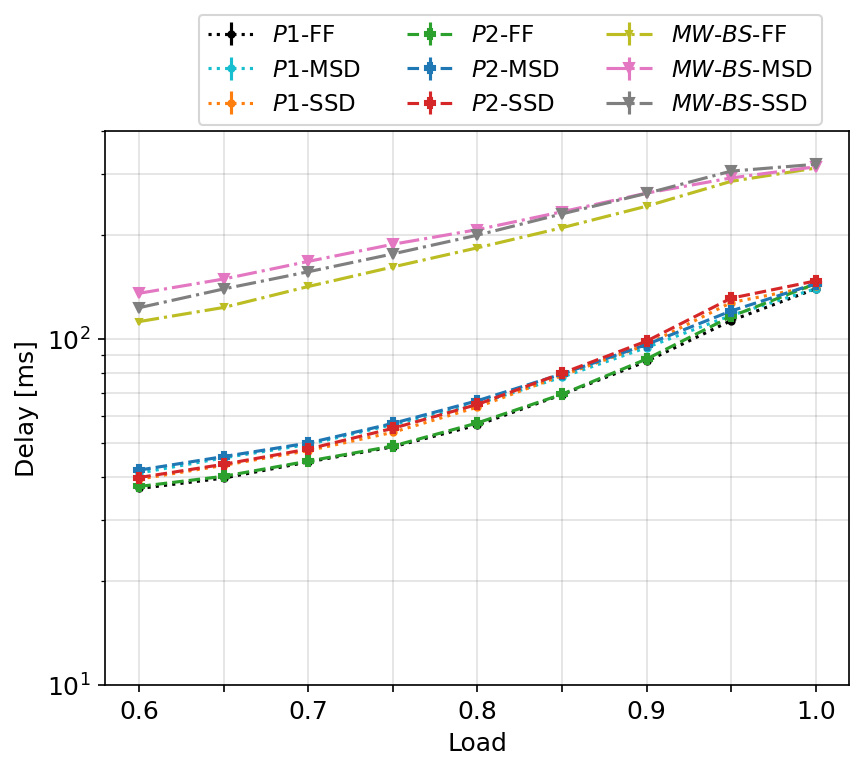}}%
\caption{Average delay produced by the \delaycritical and \ac{FL} applications. $P1$ and $P2$ mean, respectively, \FLx algorithm with \policyone and \policytwo.}
\label{fig:mean_delays}
\end{figure*}

\begin{figure*}
\centering
\subfloat[Involved clients vs. synchronization time]{\label{fig:sym}
\includegraphics[height=2.3in]{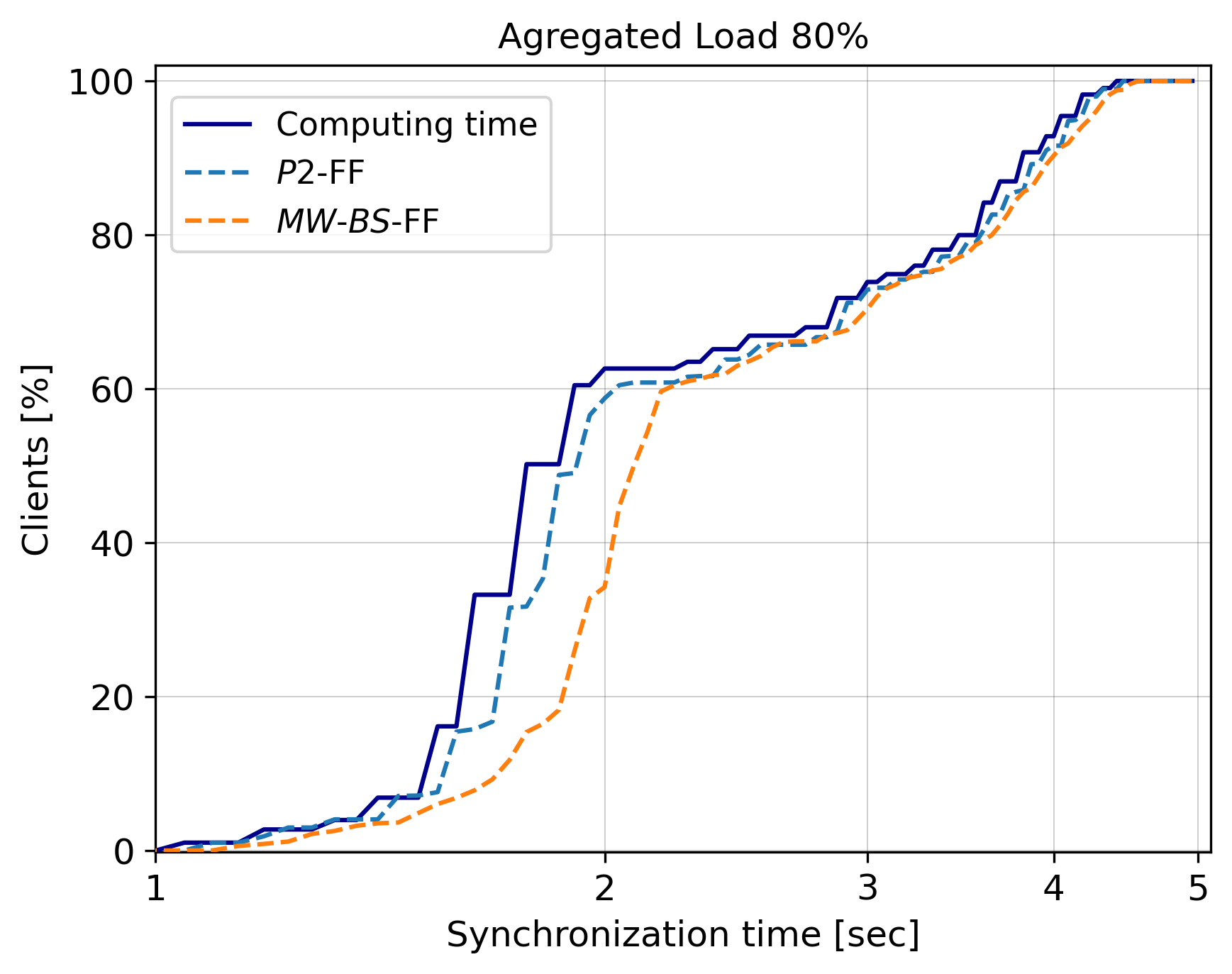}}%
\qquad
\subfloat[Learning accuracy (with 2000 rounds) vs. synchronization time]{\label{fig:acc}%
\includegraphics[height=2.3in]{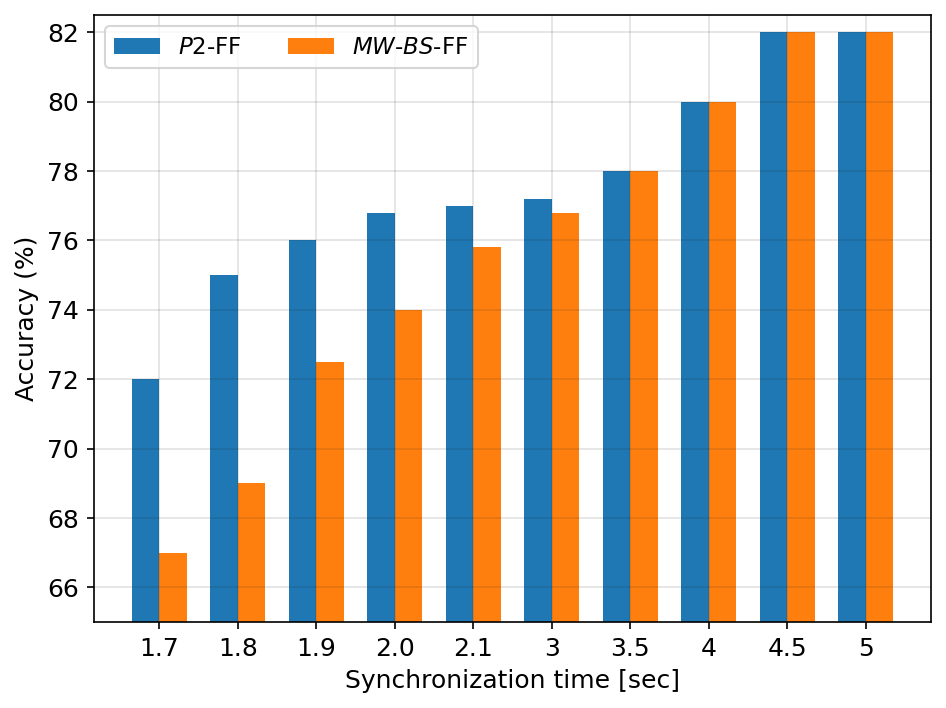}}%
\caption{Percentage of the learning accuracy and involved clients depending on the synchronization time. $P2$ means \FLx algorithm with \policytwo.}
\label{fig:results2}
\end{figure*}


\begin{figure*}%
\centering
\includegraphics[height=2.0in]{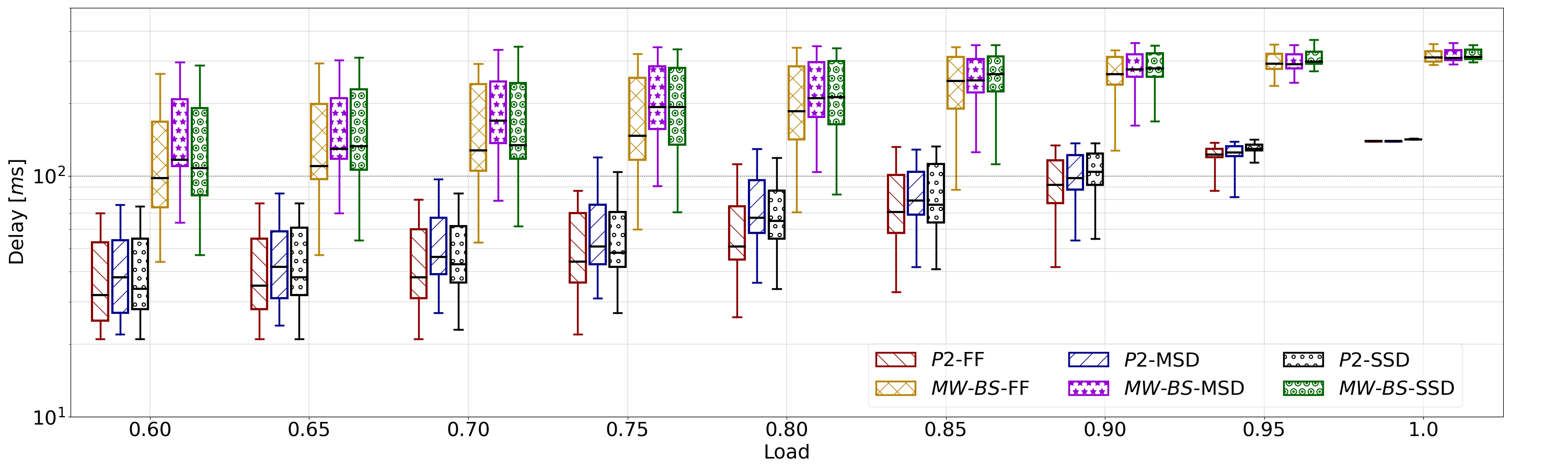}%
\caption{
Network delay performance for the \acl{FL} application. Each boxplot shows 10th, 30th, 50th, 80th, and 100th percentile of network delay for FL clients; $P2$ means \FLx algorithm with \policytwo..}
\label{fig:results}
\end{figure*}

\subsection{Simulation Results and Discussion}
\label{sec:results}

Mean delay values obtained by DWBA-FL  were  lower than \unit[80]{ms} and \unit[150]{ms} for the \ac{FL} traffic in both underloaded and overloaded conditions, respectively.
The  delay values given by \ac{BS} were at least twice as high as those given by our proposal (Fig.~\ref{fig:fl}).
This improvement is a consequence of the large windows allocated for transmissions of \ac{FL} traffic when our proposal is employed. 

Moreover, the use of the  \policytwo 
produced lower delay values for the \delaycritical traffic lower than those given by both the \policyone and the \BS algorithm (Fig.~\ref{fig:ef}).
This result occurs because the bandwidth slice is statically allocated for the \ac{FL} traffic. 
Furthermore, the strict prioritization of  \ac{FL} traffic employing the \policyone and the huge amount of traffic produced by the \ac{FL} application leads to delay-critical application to bandwidth starvation.
The mean delay of the \delaycritical traffic produced by the \policyone was from \unit[200]{us} to \unit[1000]{us} and is more than that produced by  \policyone. 
Thus, the DWBA-FL with \policytwo produce mean delay values for the \acl{FL} and \delaycritical application lower than the other algorithms.

Furthermore, the \ac{FF} policy produces a slight decrease in  delay values for both traffics than the other wavelength allocation policies (\ie SSD and MSD).
This results are a consequence of the wastage of bandwidth due to the excessive uses of guard periods and  poor multiplexing gain when employing the \ac{SSD} and \ac{MSD}, respectively.

In Fig. \ref{fig:sym}, the blue curve shows the proportion of  clients involved as a function of the computing time. It shows the minimal synchronization time per round without any communication delay.
The \BSx algorithm requires a longer synchronization time per round to produce the same percentage of the involved clients than that required by the proposed scheme with \policytwo. 
For example, it is required a synchronization time of \unit[1.9]{s} and \unit[2.1]{s} to produce a percentage of  involved clients of \unit[50]{\%} with our proposal and the \BSx algorithm, respectively.
To achieve a training accuracy of 76\% (Fig. \ref{fig:acc}), the proposed scheme can reduce 9.5\% of the training time compared to the \ac{BS} algorithm (\ie \unit[0.2]{s} less for a synchronization time of \unit[2.1]{s}), when the total traffic load is 0.8.

Fig. \ref{fig:results} shows the  network delay as a function of the  ONU offered load.
%
%
The \BSx produce delay values greater than \unit[300]{ms}, whereas, with the DWBA-FA algorithm, these values are reduced to less than  \unit[150]{ms}. 
Moreover, for \unit[80]{\%} of the clients, which is the typical  percentage of clients that produce  accuracy greater than \unit[75]{\%} (see Fig. \ref{fig:results2}), the \BSx scheme imposes  a network delay greater than \unit[200]{ms}, while the DWBA-FA imposes delay values lower than \unit[100]{ms}, under underloaded conditions (\ie load < 0.85).
In summary, DWBA-FL reduces the network delay when compared to the \BSx scheme.
This reduction in delay may decrease the number of stragglers, which in the end leads to a faster convergence and greater model accuracy.

	\section{Conclusion}
\label{sec:conclusion}

This paper has introduced a resource allocation (RA) scheme for supporting \acl{FL} applications in TWDM-\acp{EPON} networks. 
Our proposal includes a strict prioritization for  federated learning and  \delaycritical traffic, which maximizes the allocated bandwidth and reduce the delay for both types of  application.
Our proposal reduces the synchronization time without compromising the number of  involved clients. It also reduces the delay of time-critical and federated learning applications.  
Future research directions are envisioned as follows. 
Various \ac{FL} applications  coexisting on a given  network infrastructure, which have different size of the local model parameters, number of clients and synchronization time.
Mechanisms are needed to  appropriately address the \ac{QoS} provisioning for the diversity of  FL applications.
These schemes may schedule the \ac{FL} traffics based on required bandwidth but also considering the number of straggler clients, the diverse \ac{FL} packet sizes, and synchronization time.

\section*{Acknowledgments}

This work was partially sponsored by grant \#15/24494-8, São Paulo Research Foundation (FAPESP), and CNPq.

\bibliographystyle{IEEEtran}
\bibliography{IEEEabrv,CoopMOS-IPACT}

\begin{IEEEbiographynophoto}
{Oscar Ciceri} (oscar@lrc.ic.unicamp.br) completed his five-year degree in Electronics and Telecommunications Engineering at Universidad del Cauca (UNICAUCA),  Colombia,  in 2015, and his M.S. degree in Computer Science at the University of Campinas (UNICAMP),  Brazil, in 2019. He is a Ph.D. student and researcher in the Network Computing Laboratory (LRC) at that university. His research interests include Passive Optical Networks (PONs), 5G and 5G beyond networks, Machine Learning, and virtualization. His current work is on the low-latency PON technologies for 5G and 5G beyond mobile optical access systems.
\end{IEEEbiographynophoto}
\vspace{-0.6cm}
\begin{IEEEbiographynophoto}
{Carlos  Astudillo} (castudillo@lrc.ic.unicamp.br) received his bachelor's degree in Electronics and Telecommunications Engineering from the University of Cauca, Colombia in 2009, and his MSc degree in Computer Science from the University of Campinas (UNICAMP), Brazil, in 2015. Currently, he is a PhD candidate in Computer Science at the Institute of Computing (IC), UNICAMP. His master's thesis received 1st place at the 29th Theses and Dissertations Contest (CTD) of the Brazilian Computer Society (SBC), and the 2016 Best Master's Thesis Award from IC/UNICAMP. His research interests include radio resource allocation and quality of service provisioning for machine to machine communications and the Internet of Things, backhauling/fronthauling in 5G \& B5G cellular networks, and machine learning for networking.
\end{IEEEbiographynophoto}
\vspace{1.6cm}

\begin{IEEEbiographynophoto}
{Zuqing Zhu} [M’07, SM’12] (zqzhu@ieee.org) received his Ph.D. degree from the Department of Electrical and Computer Engineering, University of California, Davis, in 2007. He is currently a full professor at USTC. He has published more than 200 papers in referreed journals and conferences. He was an IEEE Communications Society Distinguished Lecturer (2018–2019) and a Senior Member of OSA.
\end{IEEEbiographynophoto}

\vspace{-15.2cm}

\begin{IEEEbiographynophoto}
{Nelson L. S. da Fonseca} [M'88, SM'01] (nfonseca@ic.unicamp.br) obtained his Ph.D. degree from the University of Southern California in 1994. He is Full Professor at the Institute of Computing, State University of Campinas (UNICAMP). He published 400+ papers and supervised 70+ graduate students. He is Senior Editor for the IEEE Communications Magazine and the IEEE Systems Journal. He is also an Associate Editor for Peer-to-Peer Networking and Applications and Computer Networks. He is past EiC of the IEEE Communications Surveys and Tutorials. He served as the IEEE ComSoc VP Publications, VP Technical and Educational Activities, and VP Member Relations.
\end{IEEEbiographynophoto}

\end{document}